\documentstyle[graphicx]{aipproc}
 
\begin{document}
\title{The Supernova-GRB Connection}
\author{P. H\"oflich$^1$, A. Khokhlov $^2$, S. Rosswog$^3$, L. Wang$^4$}
\address{
$^1$ Depart.of Astronomy, U. of Texas, Austin, TX 78681, USA\\
$^2$ Naval Research Lab, Washington DC, USA\\
$^3$ University of Leicester, Leicester LE1 7RH, UK\\
$^4$ Lawrence Berkeley Lab, 1 Cyclotron Rd, Berkeley, CA 94720, USA
}
\maketitle
\begin{abstract} 
 We discuss the possible connection between supernova  explosions (SN) and gamma-ray bursters (GRB) from
the perspective of our current understanding of SN  physics.
 Recent evidence strongly suggests that the explosion mechanism of core collapse SN
is intrinsically aspherical. Typically, a neutron star is formed.
 However, the observed properties of the expanding SN envelopes remnants make these objects very unlikely
candidates for GRBs.
Most  candidates for a GRB/SN connection seem to require the prompt or delayed formation of a black hole.
These include the collapse of very massive stars (e.g. hypernovae) and  'classical' SNe with a significant
 fallback of material over time scales of hours to days,  resulting in the collapse of the neutron star to a 
black hole.
We suggest the merger of a neutron star with a white dwarf as a subclass of thermonuclear SNe
and a potential candidate for a SN/GRB connection.
\end{abstract}

\section {Introduction}
 This volume in honor of Jan van Paradijs can be regarded as a state of the art reference on, basically, all relevant
aspects and observations for GRBs. Basically, references should be given to almost all presentations. For practical reasons,
we want to refer globally to those at this point.
 The evidence for connection between supernovae (SN) and gamma-ray bursters (GRB) include the coincidence between SN1998bw and GRB~980425,
bumps in the afterglow of some GRBs which are compatible with an underlying SN explosion,
possible identification of the Fe K line in the X-rays, and some evidence that GRBs are
related to star forming region.
 It becomes increasingly obvious that  both the mechanism causing GRBs and SNe
are intrinsically highly aspherical. 
 These evidences provide the basis for the current suggestions  that the
SN and GRB are connected.
 In particular, the direct  collapse of the central region a  very massive star to a black hole and the connection 
 with the newly found hypernovae is a very attractive option (e.g. Woosley \& MacFayden 1999, Fryer et al. 1999).
 In this scenario, the core 
of a massive star collapses directly to a black hole. Subsequently, an accretion disk is formed and a
ultra-relativistic jet is  launched. About 0.5 $M_\odot $ of radioactive Ni
is ejected. Despite the evidence,
there are are number of
unsolved problems.  GRB ~980425 was about a factor of 1000 dimmer than typical GRBs. The properties of the
optical afterglow of GRBs strongly suggest a very low density environment unlike those to be expected
in star forming regions. 
A low barion load is required despite the jet-launch in a high density star,
and there seems to be no coincidence between SNII, SNeIb/c and GRBs. From theory and based on 2-D
models, the formation of highly relativistic jets has been demonstrated to occur (Aloy et al. 2000) but the energies of the
ultra-relativistic parts of the jet are smaller
by many  orders of magnitude ($\approx 10^{-5}$) compared to the emitted $\gamma $-ray energy, and the question on
the stability of the narrow  jets has been posed. Moreover, if $^{56}Ni$ is ejected, we would expect rather
two K-shell lines due to $^{56}Ni$ and $^{56}Co$ rather than one feature as reported.
 We note that neither of the problems must be regarded
as a killer for the hyper-supernovae/GRB connection but they have to be solved.\\
 Depending on taste, we may feel close to the solution of the GRB problem, or far away.
One of the reasons is the lack of  understanding of the 'normal' core collapse SNe
 or, more precisely, the mechanism which results into the ejection of the envelope. 
 Here, we want to ask  for the current status of
core collapse SNe  and their relation to  GRBs.
 At the end, we want to discuss the merging between a neutron star and WD
as  an additional scenario which, at the same time,  may account for some of the subluminous
thermonuclear explosions (SNeIa) and some of the GRBs.

\section{Core Collapse Supernovae}

 Core collapse supernovae  are the final stages of stellar evolution in 
massive stars during which the central region collapses, forms a neutron star 
(NS) or a black hole, and the outer layers are ejected. Recent explosion scenarios assumed that
the ejection is due to energy deposition by neutrinos into the envelope but 
detailed models do not produce powerful explosions and, in most cases, do not trigger the ejection of the envelope at all
 (e.g. Yamada et al. 1999, Ramp \& Janka 2000,
Mezzacappa et al. 2001).

 There is new and mounting evidence
for an asphericity and, in particular, for axial symmetry in several SNe which
may be hard to reconcile within the spherical picture.
  This evidence includes the observed high polarization and its variation with time (e.g. Wang et al. 2001,
Leonard et al. 2000),
pulsar kicks (Strom et al. 1995), high velocity iron-group and intermediate-mass element material
observed in remnants (Lucy 1987, Tueller et al. 1991),  direct observations of the debris of SN1987A (Wang et al. 2001,
H\"oflich, Khokhlov \& Wang 2001), etc.
 To be in agreement with the observations, any successful mechanism must invoke
some sort of axial symmetry for the explosion.
 As a limiting case, we consider jet-induced/dominated explosions of "classical" 
core collapse SNe. The discovery of magnetars (Kouveliotou et al. 1998, Duncan \& Thomson 1992)
 revived the idea that a MHD-jet with appropriate properties may be formed at the NS or BH (LeBlanc \& Wilson 1970, Symbalisty 1984).
 Our study is  based on detailed 3-D hydrodynamical and radiation transport models (Khokhlov 1998, H\"oflich et al. 1998, 2001).
 We demonstrate the influence of the jet properties and 
of progenitor structure on the final density, chemical structure and the fallback of material.

\subsection{Results for Jet-Induced Supernovae}

\noindent
 {\bf The Setup:}
   The  computational domain is a cube of size $L$    
with a spherical star of radius $R_{\rm star}$ and mass $M_{\rm star}$
 placed in the center.
 The innermost part with mass $M_{\rm core} \simeq 1.6 
M_{\odot}$ and radius  $R_{\rm core} = 4.5    \times  10^8$~cm, consisting
of Fe and Si, is assumed to have collapsed on a timescale much faster
than the outer, lower-density material. It is removed and replaced by a
point gravitational source with mass $ M_{\rm core}$ representing the
newly formed neutron star.  The remaining mass of the envelope $M_{env}$
is mapped onto the
computational  domain. 
 At two polar locations where the jets
are initiated at $R_{core}$, we impose an inflow with velocity $v_j$ and a density
$\rho_j$.   
 At  $R_{\rm core}$,
the jet density and pressure are the same as those of the background
material.
For the first 0.5~s, the jet velocity at $R_{\rm core}$ is kept
constant at $v_j$.
  After 0.5~s, the
velocity of the jets at $R_{\rm core}$ was gradually decreased to zero
at approximately 2~s. The total energy of the jets is $E_j$.
 These parameters are
consistent with, but somewhat less than, those of the LeBlanc-Wilson model.

\noindent
{\bf The reference model:} As a baseline case, we consider a jet-induced explosion in a helium
star. 
  Jet propagation inside the star is shown in
Fig. \ref{jet1}.
 As the jets move outwards, they remain collimated and do not  develop much internal
structure. A bow shock forms at the head of the jet and spreads in all
directions, roughly cylindrically around each jet.   
  The jet-engine has been switched off after about 2.5 seconds,
the material of the bow shock continues to propagate through the star.                    
 The stellar material is shocked by the bow shock. Mach shocks
travels towards the equator resulting in a redistribution of the 
energy. The opening angle of the jet depends on the ratio between the
velocity of the bow shock to the speed of sound. For a given star,    
 this angle determines the efficiency of the deposition of the
 jet energy into the stellar envelope. Here, the
efficiency of the energy deposition is about 40 \%, and
the final asymmetry of the envelope is about two.

\noindent
 {\bf Influence of the  jet properties:}
 Fig. \ref{jet2} shows two examples of an explosion with
 a low and a very high jet velocity compared to the baseline case (Fig. \ref{jet1}).
  Fig. \ref{jet2}  demonstrates the influence of the jet velocity on the opening
 angle of the jet and, consequently,
 on the efficiency of the energy deposition. For the low velocity jet,
the jet engine is switched off long before
the jet penetrates the stellar envelope. Almost all of the energy of the
jet goes into the stellar explosion. On a contrary, the  fast jet (61,000 km/sec)
 triggers only a weak explosion of 0.9 foe although its 
 total energy was $\approx 10 foe $.

\noindent
{\bf Influence of the progenitor:}
 For a very extended star, as  in  case of 'normal'  Type II Supernovae,
the bow shock of a low velocity jet stalls within the envelope, and
 the entire jet energy is used to trigger the ejection of the stellar envelope. In
our example (Fig. \ref{model}),
the jet material penetrates the helium core at about 100 seconds.
 After about 250 seconds the material of the jet  stalls within the
hydrogen rich envelope and
 after passing about 5 solar masses in the radial
mass scale of the spherical progenitor.
 At this time, the isobars are almost spherical,  and an almost 
spherical shock front travels outwards. Consequently, 
strong asphericities are 
limited to the inner regions.
 After about 385 seconds,  we stopped the 3-D run and remaped the  outer layers into 
1-D structure, and followed the further evolution in 1-D.
 After about 1.8 $10^ 4$ seconds, the shock front reaches 
the surface. After about 3 days, the envelope expands homologously.
The region where the jet material stalled, expands at velocities of about
4500 km/sec.
 
\noindent
{\bf Fallback:}
 Jet-induced SN  have very different characteristics with
respect to fallback of material and the innermost structure.
 In 1-D calculations and for stars with
main sequence masses of less than  20 $M_\odot$ and explosion
energies in excess of 1 foe, the fallback of material remains less than
1.E-2 to 1.E-3 $M_\odot$ and an inner, low density cavity is formed 
with an outer edge of $^{56}Ni$. For explosion energies between 1 and 2 foe,
 the outer edge of the cavity expands typically with
velocities of about 700 to 1500 km/sec
(e.g. Woosley 1997,  H\"oflich et al. 2000).
 In contrast, we find strong, continuous fallback of     
$\approx 0.2  M_\odot$ in our 3-D models,
and no lower limit for the velocity of  the expanding material (Fig. 4 of Khokhlov \& H\"oflich 2001).
 This significant amount of fallback must have important consequences for the
secondary formation of a black hole.
 The exact amount and time scales for the 
final accretion on the NS will depend sensitively on the
rotation and momentum transport.

\noindent
{\bf Chemical Structure:}
 The final chemical profiles of elements 
formed during the stellar evolution  such as He, C, O and Si
are 'butterfly-shaped' whereas the jet material fills
an inner, conic structure (Fig. \ref{model}, upper, middle panel).
 
 The
composition of the jets must reflect the composition of the innermost
 parts of the star, and should contain heavy 
and intermediate-mass elements, freshly synthesized material such
as $^{56}Ni$ and, maybe, r-process elements because, in our examples,
the entropy at the bow shock region of the jet was as high as a few hundred,
a crucial condition for a successful r-process in a high-$Y_e$ environment.
In any case,
during the explosion, the jets bring heavy and intermediate mass elements
into the outer H-rich  layers.

\noindent
{\bf Radiation Transport Effects:}
 For the compact progenitors of SNe~Ib/c, the final departures of the 
iso-density contours from sphericity  are typically a factor of two. This
 will produce a linear polarization of about
2 to 3 \% (H\"oflich et al. 1995) consistent with the values observed for SNe~Ib/c.
 In case of a red supergiant, i.e. SNe~II,
the   asphericity is restricted to the inner layers of the H-rich envelope. There
the iso-densities show an axis ratio  of up to $\approx$ 1.3.
 The intermediate and outer H-rich layers remain spherical.
Thus, even within the picture of jet-induced explosion, the latter effect alone
cannot (!) account for the high polarization produced in the intermediate H-rich layers
of core-collapse SN with a massive envelope such as  SN~1999em where asymmetric ionization
by $\gamma $ photons for an aspherical chemical distribution of $^{56}Ni$
 is crucial (see Fig. \ref{model} and
H\"oflich et al. 2001).

\noindent
{\bf Conclusions:}
Our calculations show that low velocity, massive jets can explain the observations of core collapse  SNe.
 These kind of jets, however, will not produce  typical
GRB energies though $\gamma$-rays with total energies of $\approx 10^48$ may be produced at the jet-breakout in
case of
the explosion of a SNIc (Khokhlov et al. 1998) and, within this picture, events such as
 SN1998bw/GRB~980425 may be understood. However, the typical properties of the jets make  supernovae
with neutron star  remnants very  unlikely candidates to produce  GRBs within the same scenario.
All potential candidates for a GRB/SN connection seem to require the prompt (MacFayden et al. 2001)
 or delayed formation of a black hole. Whether this is sufficient, remains to be seen.

\section{An alternative scenario:\\ neutron star - white dwarf mergers}
 In this section, we want to propose an alternative scenario for SN-GRB-connection:
the merger of a  CO white dwarf (WD) and a NS.
These systems are expected to be formed at a relatively high rate, $10^{-4}-10^{-5}$ per year and galaxy
and a large fraction of them will merge within a Hubble time via the emission of gravitational waves, therefore
making an important contribution to the population of potential GRB progenitor systems (Davies et al., 2001).
In some of these systems, e.g. J1141-6546, the WD seems to have formed {\em before} the NS and therefore the 
time scale to form these systems is set by the neutron star formation time scale. Once formed, 5\% (20\%, 50\%)
of the binaries will merge within 10$^{6}$ (10$^{7}$, 10$^{8}$) years, reconciling this scenario with 
the observation that (at least some) GRBs seem to occur in star forming regions.\\
The basic phenomenology of the scenario is as follows: the WD transfers mass to the NS, which subsequently 
settles into an accretion disk around the NS. Once enough material is accreted, the NS may collapse
to a black hole (BH). As we will see below, this event may show
properties similar to thermonuclear SNe, i.e. SNe~Ia  and, at the same
time, may contribute to the phenomenon called GRBs. Though, this event cannot be the
main contributor to either SNe~Ia or GRBs because of the rate and properties of typical
SNe~Ia or of GRBs (Piran, this volume).

{\bf Simulation:} In the following example, the results are based on a calculation using a smoothed particle hydrodynamics
code, in which self-gravity is calculated using a binary-tree, the equation of state accounts for the nuclei, 
electron-positron pairs and photons. For further details we refer to Rosswog et al.
(2001, 2000, 1999). We have followed the final stages of the merging process between a neutron star and a C/O 
white dwarf of 0.7 $M_\odot$ and a radius of 7000 km. The neutron star is modelled as a point mass
of 1.4 $M_\odot$, the WD is resolved with $\sim 110 \; 000$ particles. To model all the details of the mass 
transfer from the WD to the NS from the onset 
until the final WD-disruption is an almost impossible task: the initial mass transfer happens with rates
 too tiny to be
resolved accurately in a full 3D calculation and, even worse, happens on time scales much too long to be followed
by an explicit code whose time step is set by the sound velocity within the WD. Therefore the initial separation
has to  be chosen as a compromise between accuracy and feasibility, we chose 22,000 km in order to see the onset 
of particles being transferred to the NS within the first orbit. Therefore we neglect the
existence of a low-mass disk which may exist before the tidal disruption sets in, but for the current, exploratory
purposes this treatment is sufficent.
The evolution of the system can be seen in Fig. \ref{merger}. 
Within one orbital revolution mass transfer towards the NS sets in and leads to the formation of an accretion disk.
This disk is fed continuoulsy from the WD for roughly four orbital periods before, at about 150 s, tidal disruption
sets in. \\ 
{\bf SN:} The corresponding temperature and density
distributions are shown as a function of the distance to the neutron star at 150 s. At this time, typical densities
and temperatures at the central region of the  disk ($r \leq 5000...7000$ km) are 1 to 3 $\times 10^7$ g/cm$^3$ and in
excess of 1.3$\times 10^9$ K, respectively. These conditions are remarkable similar to those found in
SNeIa. Under these conditions, the nuclear burning time scales are of the order of 0.1 to 1 sec,
i.e. comparable or shorter than the hydrodynamic time scales. A nuclear detonation front will form and burn 
parts of the envelope. Assuming a time scale of 1 sec, all material with $log (\rho)$ larger than 6, 5.6  
and 5 g/cm$^{3}$ will 
undergo complete Si, explosive O and C burning, respectively. The corresponding  products are  mainly $^{56}Ni$, 
$^{28}Si$ and $^{16}O$, respectively. In our example, only $\approx 0.1 M_\odot$ of $^{56} Ni$
can be expected to be produced, and about 0.4 to  0.5 $M_\odot $ are ejected during the thermonuclear explosion.
Detailed calculations for the further hydrodynamical evolution and light curves are underway using our 
radiation hydro code including a detailed nuclear network. This scenario will not be
able to account for a typical SNe~Ia but it may contribute to  the subclass of very underluminous SNe~Ia,
and it may be able to account for the bumps in some afterglows and the iron K line in the X-ray spectra 
observed in some GRBs. \\
{\bf GRB:} At the end of the simulation roughly 1.9 $M_\odot$ are contained in the NS+disk-system, 
around 0.2 $M_\odot$ form the tidal tail that moves outwards. If we assume that, eventually,  the SN-explosion forms a
black-hole and that $\sim 0.1 M_\odot $ of material in the tidal tail survives the explosion while orbiting 
around the BH, we are left with the 'standard' GRB-model, BH + debris, and 
 an energy reservoir of a few times 10$^{52}$ erg, enough for the estimated, typical GRB-energy of $5 \times
10^{50}$ erg (Frail et al. 2001). The GRB may occur when the surviving debris falls back
towards the BH, the delay between the SN-explosion and the GRB can be expected to be of the order
$\tau_{delay} \approx \tau_{orbit} \approx 2 \mbox{ h} \left(\frac{a}{100 R_{WD}}\right)^{3/2}$,
where $a$ is the semi-major axis of the debris orbit, but this delay is very sensitive to the 
details of the individual merger process.\\
The current study should be seen as  pilot project to investigate the potential
of this scenario. Many aspects have not yet been explored in full. For example, we did not investigate the 
details of the WD-spins, the ignition process as a function of time, the details of the accretion on the 
neutron star and so on. These topics are subject to future investigations.

\noindent
 \bf{Acknowledgement}:{\sl S.R. acknowledges the use of the UK Astrophysical Fluids Facility (UKAFF)
and of the University of Leicester Mathematical Modelling Centre's
supercomputer (HEX) for the SPH calculations  reported above. In parts, this work                                
has been supported  by  NASA Grant NAG5-7937.}

\newpage
{\bf Figure Captions:}
\begin{figure}[ht]
\caption {
 Observational evidence for asphericity in core collapse supernovae.
The HST image of SN~1987A (left panel) shows the inner debris of the SN-ejecta with an axis ratio of $\approx 2$ and the ring.
 Note that the inner ring
has been formed during the stellar evolution  about 10,000 years before the explosion.
The right panel shows the evolution with time of the linear polarization $P$ in the plateau
 SN~1999em. Although $P$ increased with time,  the
 polarization angle remained constant with time and wavelength indicating a common
 axis  of symmetry in the expanding envelope (Leonard et al. 2001, Wang et al. 2001, H\"oflich et al. 2001).
}
\label{obs}
\vskip -0.05truecm
\end{figure}

\begin{figure}[ht]
\caption {Logarithm of the density structure 
as a function of time for a helium core.
The total mass of the ejecta is 2.6 $M_\odot$.
The initial radius, velocity  and density of the jet were taken to 1200 km,
32,000 km/sec and $6.5 10^5 g/cm^3$, respectively. The shown domains 7.9,
9.0, 36 and 45 $\times 10^9 cm $.
The total energy is about 9 $10^{50}$ erg. After about 4.5 seconds, the 
jet penetrates the star. The energy deposited in the stellar envelope by the jet
is about 4 $10^{50}$ erg, and the final asymmetry is of the order of two.}
\label{jet1}
\vskip -0.03cm
\end{figure}

\begin{figure}[ht]
\caption {Same as Fig. \ref{jet1} ($0.5 \leq log(\rho)\leq 5.7 $)
 but for a jet velocity of 61,000 km/sec and
a total energy of 10 foe at $\approx 1.9 sec$ (left), and 
 11,000 km/sec and a total energy of 0.6 foe (right).  The size of the
presented domains are 5 (left) and 2 $10^{10} cm$ (right), respectively.
 For the high velocity jet,
 most of the energy is carried away by the
jet. Only 0.9 foe are deposited in the expanding envelope.
 In case of a low velocity jet, the bow-shock still propagates through
the star after the jet is switched off,                   and
the entire jet energy is deposited in the expanding envelope.}
\vskip -0.03cm
\label{jet2}
\end{figure}

 \begin{figure}[t]
 \caption {Polarization produced by an aspherical, chemical distribution
  for a SN~IIp model with $15 M_\odot$ and an explosion energy $E_{exp} = 2 \times 10^{51}erg$.
   This model resembles the extreme plateau SN 1999em.
  The initial density profile is given for a star at the final stage of stellar evolution
  for metalicities Z of 0.02, 0.001 and 0 (models 15a, 15b, 15z, upper left panel,
   from H\"oflich et al. 2000 \& Chieffi et al. 2001). The model for the  RSG,
   15a, has been used to calculation of the jet-induced explosion).
    In the upper, middle panel, the chemical distribution of He
    is given at 250 sec for the He-rich layers after the jet material has stalled.
  The color-codes white, yellow, green, blue and red correspond to He mass fractions of
  0., 0.18, 0.36, 0.72, and  1., respectively.
   The subsequent explosion has been followed in 1-D up to the phase of homologous
  expansion. In the upper, right panel, the density distribution is given at about 5 days
  after the explosion.  The steep gradients in the density in the upper right and left panels
  are located at the interface between the He-core and the H-mantel.
   In the lower, left panel,
  the resulting bolometric LCs are given for a our $E_{exp}=$
   $2 \times 10^{51}erg$ (dotted line) and, for comparison, for $1\times 10^{51}erg$, respectively.
   Based on full 3-D calculations for the radiation \& $\gamma $-ray transport,
  we have calculated the location of the recombination front as a function of
  time. The resulting shape of the photosphere is always prolate.
   The corresponding axis ratio and the  polarization seen from the equator are shown
  (lower, right panel).
  Note the strong increase of the asphericity after the onset of the recombination phase between
  day 30 to 40  (see also SN~1999em in Fig. \ref{obs}).
 }
\vskip -0.03cm
\label{model}
\end{figure}

\begin{figure}[ht]
\vskip -0.03cm
\vskip -0.03cm
\caption { Tidal disruption of a C/O WD with 0.7 $M_\odot$ in a WD/NS system (see text), and the
formation of an accretion disc around the neutron star at t=0, 46.6, 95.9 \& 150.7 seconds.
 The different panels show the distribution of material
projected into the orbital plane. The axis are given in units of $10^9 cm$.
The red circle indicates the location of the neutron star with $1.4 M_\odot$.
}
\vskip -0.03cm
\label{merger}
\end{figure}

\begin{figure}[ht]
\vskip -0.03cm
\caption {
Density (left) and temperature (middle) as a function of radial distance from the NS at t=150s. The highest
densities and temperatures are achieved in the orbital plane. In the right panel,
the enclosed amount of C/O as a function of radial distance from the neutron
star at t=80, 148 and 175 seconds.}
\label{merger2}
\vskip -0.04cm
\end{figure}
 
\end{document}